\documentclass[RNAAS]{aastex63}

\graphicspath{{./}{figures/}}
\begin{document}

\title{Prospects for the In Situ detection of Comet C/2019 Y4 ATLAS by Solar Orbiter}


\correspondingauthor{Geraint H. Jones}
\email{g.h.jones@ucl.ac.uk}

\author[0000-0002-5859-1136]{Geraint H. Jones}
\affiliation{Mullard Space Science Laboratory, University College London, Holmbury St. Mary, Dorking, Surrey, RH5 6NT, UK}
\affiliation{Centre for Planetary Sciences at UCL/Birkbeck, London, UK}

\author[0000-0001-8788-5494]{Qasim Afghan}
\affiliation{Mullard Space Science Laboratory, University College London, Holmbury St. Mary, Dorking, Surrey, RH5 6NT, UK}
\affiliation{Centre for Planetary Sciences at UCL/Birkbeck, London, UK}

\author{Oliver Price}
\affiliation{Mullard Space Science Laboratory, University College London, Holmbury St. Mary, Dorking, Surrey, RH5 6NT, UK}
\affiliation{Centre for Planetary Sciences at UCL/Birkbeck, London, UK}



\begin{abstract}
The European Space Agency's Solar Orbiter spacecraft will pass approximately downstream of the position of comet C/2019 Y4 (ATLAS) in late May and early June 2020. We predict that the spacecraft may encounter the comet's ion tail around 2020 May 31 - June 1, and that the comet's dust tail may be crossed on 2020 June 6. We outline the solar wind features and dust grain collisions that the spacecraft's instruments may detect when crossing the comet's two tails. Solar Orbiter will also pass close to the orbital path of C/2020 F8 (SWAN) on 2020 May 22, but we believe that it is unlikely to detect any material associated with that comet.
\end{abstract}

\section{} 

Comet C/2019 Y4 (ATLAS), discovered on 2019 December 28, will reach its 0.253~au perihelion early on 2020 May 31, coincidentally less than 6 hours before the comet crosses the ecliptic plane. Its perihelion distance is well inside the orbit of Mercury, meeting the definition of a Near-Sun Comet \citep{jones2018}. During the first months of 2020, Comet ATLAS exhibited an activity increase suggesting that it could become very active when closest to the Sun.
However, around 70 days pre-perihelion, the comet began to fade, and its nucleus underwent fragmentation \citep{hui2020observations}. At the time of this writing, the fragmented object is still active, and may yet survive to reach perihelion, albeit with lower gas and dust production rates than anticipated by many.

The European Space Agency's {\em Solar Orbiter} spacecraft was launched on 2020 February 10, and is equipped with a range of remote and {\em in situ} instruments to observe the Sun and inner heliosphere, e.g. \citet{Walsh_2020}, and references therein. In late May and early June 2020, the spacecraft will be approximately downstream of Comet ATLAS in the solar wind, potentially allowing the {\em in situ} detection by the spacecraft of the comet's ion and/or dust tails.
Serendipitous ion tail crossings by spacecraft are known to have occurred on several occasions, e.g. \citet{jones2000}, \citet{neugebauer07}. For such a traversal to take place, a comet needs to cross relatively close to the Sun-spacecraft line within a relatively short timeframe defined by the time taken for the solar wind, flowing at several hundred km~s$^{-1}$, to be able to carry ions from the comet's coma to the moving spacecraft. In general, the more productive the comet, the larger this miss distance, or impact parameter, needs to be for the spacecraft to detect the presence of an ion tail. 

For typical equatorial solar wind speeds of 300 -- 500 km$^{-1}$ during this phase of the solar activity cycle, we estimate that a crossing of Comet ATLAS's ion tail by \emph{Solar Orbiter} is most likely around 2020 May 31 -- June 1. Assuming a radial flow, solar wind particles detectable at the spacecraft will have approached to within $\sim$(6.7 -- 7.8)$\times$10$^{6}$~km of the nucleus. As seen from the spacecraft, Comet ATLAS will not approach close enough to the Sun to be observable by the spacecraft's remote sensing instruments.
If the comet has a high enough production rate, its ion tail could be detectable at \emph{Solar Orbiter} through the presence of pickup ions by instruments such as the Solar Wind Plasma Analyzer, SWA, and/or the presence of draping signatures in the heliospheric magnetic field by the magnetometer, MAG. Although the solar wind on average flows directly away from the Sun, deviations from this radial flow can take place, as evidenced by the \emph{Ulysses} spacecraft's crossing of the ion tail of C/1999~T1 (McNaught-Hartley) \citep{gloeckler2004} when 27.6$^{\circ}$ from the comet's orbital plane. Such large non-radial flows are occasionally observed in the inner heliosphere, e.g. \citet{owens2004}, widening the date range for a potential ion tail crossing.

\begin{figure}[h!]
\begin{center}
\includegraphics[scale=0.45,angle=0]{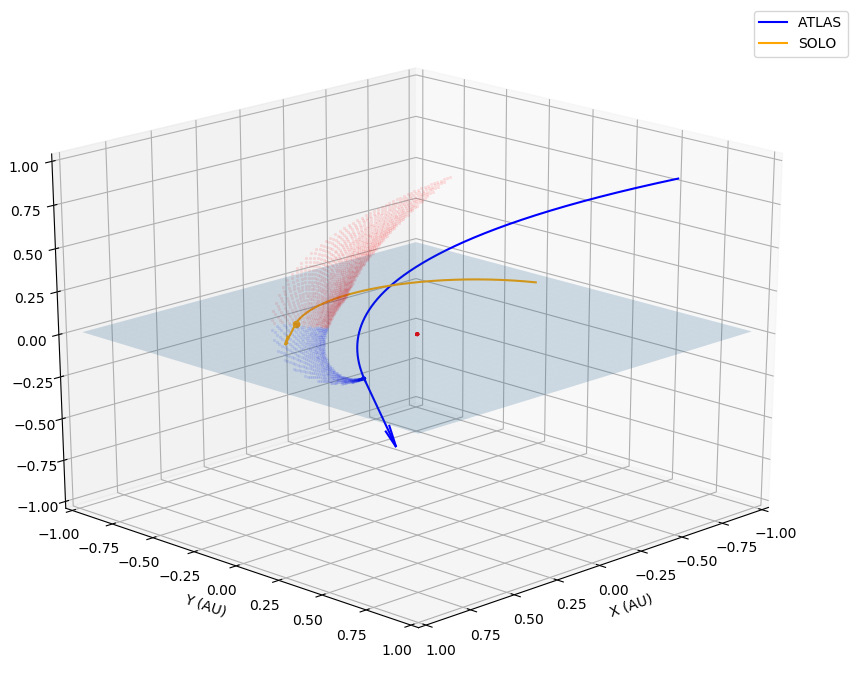}
\caption{Scale figure showing the relative paths and positions of spacecraft (orange), Comet ATLAS (blue) and its dust tail at the time on 2020 June 6 when {\em Solar Orbiter} will cross Comet ATLAS's orbital plane. The horizontal plane depicts the ecliptic plane, with the Sun shown as a red dot. Modelled dust tail particles are red to the north of the ecliptic, and blue to the south. \label{fig:1}}
\end{center}
\end{figure}

\emph{Solar Orbiter} will cross the comet's orbital plane on 2020 June 6, at around 11:40~UT (Figure~\ref{fig:1}). If the comet is active enough in the period leading up to this time, particles in its dust tail could be encountered by the spacecraft.
Using the classical dust tail modelling of \citet{Finson1968}, it is possible to constrain the grain characteristics of the dust tail section that will be traversed by \emph{Solar Orbiter}. Assuming no fragmentation of dust after release from the nucleus, the dust encountered by the spacecraft would have left the comet's nucleus around 17 days earlier, on May 20, with a $\beta$ parameter $\sim$1.2 (Figure~\ref{fig:1}).
Dust impacts on the spacecraft could be detectable by the Radio and Plasma Waves experiment, the potential signatures of which have been described by \citet{mann2019}. 

Interplanetary Field Enhancements, IFEs, are another solar wind phenomenon that may be observed around the time of the orbital plane crossing. These uncommon events are characterized by a heliospheric magnetic field magnitude increase, often with a thorn-shaped profile, and are usually accompanied by a discontinuity in the magnetic field direction. The first reported event of this kind \citep{russell1983} has been followed by several surveys and case studies, e.g. \citet{russell1984, jones2003}. Although the exact cause of these signatures is arguably still to be widely accepted, it appears clear from several associations between IFEs and cometary orbital plane crossings, e.g. \citet{lai2015,jones2003a}, that they are associated with interactions between the solar wind and dust trails associated with some comets and asteroids. We therefore suggest that IFEs should be searched for in \emph{Solar Orbiter} magnetometer data gathered on and around June 6.

We also mention in passing the fact that \emph{Solar Orbiter} will, on 2020 May 22, when at 0.60~au from the Sun, cross the orbital plane of a second comet, C/2020~F8 (SWAN). That comet will, on May 10, have passed slightly antisunward of the spacecraft's May 22 position, at 0.61~au from the Sun. An ion tail crossing appears impossible, and the chances of any dust associated with that comet being encountered by the spacecraft also appear to be extremely low.

To the authors' knowledge, if \emph{Solar Orbiter} instruments detect material from Comet ATLAS, it will be the first predicted serendipitous comet tail crossing by an active spacecraft carrying appropriate instrumentation for the detection of cometary material.

\acknowledgments

QA and OP have both been supported by UK Science \& Technology Facilities Council Ph.D. studentships. Ephemeris data for both comet and spacecraft provided via the JPL Horizons system at https://ssd.jpl.nasa.gov/?horizons ; \citet{giorgini1997}.

\bibliographystyle{aasjournal}

\begin{thebibliography}{}
\expandafter\ifx\csname natexlab\endcsname\relax\def\natexlab#1{#1}\fi
\providecommand{\url}[1]{\href{#1}{#1}}
\providecommand{\dodoi}[1]{doi:~\href{http://doi.org/#1}{\nolinkurl{#1}}}
\providecommand{\doeprint}[1]{\href{http://ascl.net/#1}{\nolinkurl{http://ascl.net/#1}}}
\providecommand{\doarXiv}[1]{\href{https://arxiv.org/abs/#1}{\nolinkurl{https://arxiv.org/abs/#1}}}

\bibitem[{Finson \& Probstein(1968)}]{Finson1968}
Finson, M., \& Probstein, R. 1968, The Astrophysical Journal, 154, 327,
  \dodoi{10.1086/149761}

\bibitem[{{Giorgini} {et~al.}(1997){Giorgini}, {Yeomans}, {Chamberlin},
  {Chodas}, {Jacobson}, {Keesey}, {Lieske}, {Ostro}, {Standish}, \&
  {Wimberly}}]{giorgini1997}
{Giorgini}, J.~D., {Yeomans}, D.~K., {Chamberlin}, A.~B., {et~al.} 1997, in
  \baas, Vol.~28, 1099

\bibitem[{{Gloeckler} {et~al.}(2004){Gloeckler}, {Allegrini}, {Elliott},
  {McComas}, {Schwadron}, {Geiss}, {von Steiger}, \& {Jones}}]{gloeckler2004}
{Gloeckler}, G., {Allegrini}, F., {Elliott}, H.~A., {et~al.} 2004, \apjl, 604,
  L121, \dodoi{10.1086/383524}

\bibitem[{Hui \& Ye(2020)}]{hui2020observations}
Hui, M.-T., \& Ye, Q.-Z. 2020.
\newblock \doarXiv{2004.10990}

\bibitem[{{Jones} {et~al.}(2000){Jones}, {Balogh}, \& {Horbury}}]{jones2000}
{Jones}, G.~H., {Balogh}, A., \& {Horbury}, T.~S. 2000, \nat, 404, 574

\bibitem[{{Jones} {et~al.}(2003{\natexlab{a}}){Jones}, {Balogh}, {McComas}, \&
  {MacDowall}}]{jones2003}
{Jones}, G.~H., {Balogh}, A., {McComas}, D.~J., \& {MacDowall}, R.~J.
  2003{\natexlab{a}}, \icarus, 166, 297, \dodoi{10.1016/j.icarus.2003.09.007}

\bibitem[{{Jones} {et~al.}(2003{\natexlab{b}}){Jones}, {Balogh}, {Russell}, \&
  {Dougherty}}]{jones2003a}
{Jones}, G.~H., {Balogh}, A., {Russell}, C.~T., \& {Dougherty}, M.~K.
  2003{\natexlab{b}}, \apjl, 597, L61, \dodoi{10.1086/379750}

\bibitem[{{Jones} {et~al.}(2018){Jones}, {Knight}, {Battams}, {Boice}, {Brown},
  {Giordano}, {Raymond}, {Snodgrass}, {Steckloff}, {Weissman}, {Fitzsimmons},
  {Lisse}, {Opitom}, {Birkett}, {Bzowski}, {Decock}, {Mann}, {Ramanjooloo}, \&
  {McCauley}}]{jones2018}
{Jones}, G.~H., {Knight}, M.~M., {Battams}, K., {et~al.} 2018, \ssr, 214, 20,
  \dodoi{10.1007/s11214-017-0446-5}

\bibitem[{{Lai} {et~al.}(2015){Lai}, {Russell}, {Jia}, {Wei}, \&
  {Angelopoulos}}]{lai2015}
{Lai}, H.~R., {Russell}, C.~T., {Jia}, Y.~D., {Wei}, H.~Y., \& {Angelopoulos},
  V. 2015, \grl, 42, 1640, \dodoi{10.1002/2015GL063302}

\bibitem[{{Mann} {et~al.}(2019){Mann}, {Nouz{\'a}k}, {Vaverka}, {Antonsen},
  {Fredriksen}, {Issautier}, {Malaspina}, {Meyer-Vernet}, {Pavl{\r{u}}},
  {Sternovsky}, {Stude}, {Ye}, \& {Zaslavsky}}]{mann2019}
{Mann}, I., {Nouz{\'a}k}, L., {Vaverka}, J., {et~al.} 2019, Annales
  Geophysicae, 37, 1121, \dodoi{10.5194/angeo-37-1121-2019}

\bibitem[{{Neugebauer} {et~al.}(2007){Neugebauer}, {Gloeckler}, {Gosling},
  {Rees}, {Skoug}, {Goldstein}, {Armstrong}, {Combi}, {M{\"a}kinen}, {McComas},
  {von Steiger}, {Zurbuchen}, {Smith}, {Geiss}, \& {Lanzerotti}}]{neugebauer07}
{Neugebauer}, M., {Gloeckler}, G., {Gosling}, J.~T., {et~al.} 2007, \apj, 667,
  1262, \dodoi{10.1086/521019}

\bibitem[{{Owens} \& {Cargill}(2004)}]{owens2004}
{Owens}, M., \& {Cargill}, P. 2004, Annales Geophysicae, 22, 4397,
  \dodoi{10.5194/angeo-22-4397-2004}

\bibitem[{{Russell} {et~al.}(1984){Russell}, {Arghavani}, \&
  {Luhmann}}]{russell1984}
{Russell}, C.~T., {Arghavani}, M.~R., \& {Luhmann}, J.~G. 1984, \icarus, 60,
  332, \dodoi{10.1016/0019-1035(84)90194-5}

\bibitem[{{Russell} {et~al.}(1983){Russell}, {Luhmann}, {Barnes}, {Mihalov}, \&
  {Elphic}}]{russell1983}
{Russell}, C.~T., {Luhmann}, J.~G., {Barnes}, A., {Mihalov}, J.~D., \&
  {Elphic}, R.~C. 1983, \nat, 305, 612, \dodoi{10.1038/305612a0}

\bibitem[{Walsh {et~al.}(2020)Walsh, Horbury, Maksimovic, Owen,
  Rodri{\'}guez-Pacheco, Wimmer-Schweingruber, \& and}]{Walsh_2020}
Walsh, A.~P., Horbury, T.~S., Maksimovic, M., {et~al.} 2020, Astronomy {\&}
  Astrophysics, \dodoi{10.1051/0004-6361/201936894}

\end{thebibliography}

\end{document}